\documentclass[]{article}

\usepackage[T2A]{fontenc}
\usepackage[tbtags]{amsmath}
\usepackage{amsfonts,amssymb}

\usepackage{my}

\theoremstyle{plain}
\newtheorem{teor}{Theorem}

\newtheorem{lem}{Lemma}
\newtheorem{cor}{Corollary}

\theoremstyle{definition}
\newtheorem{proof}{Proof}

\newtheorem{defin}{Definition}

\newcommand{\R}{{\mathbb R}}

\newcommand{\oC}{{\mathbb C}}
\newcommand{\di}{{\rm d}}

\begin{document}

\author{A.~G.~Smirnov\Addressmark[1]}

\Addresstext[1]{ Tamm Theory Department, Lebedev Physical
Institute, RAS, Moscow, Russia, e-mail: smirnov@lpi.ru. }

\title{Localization properties of highly singular generalized functions}

\markboth{A.~G.~Smirnov}{Localization properties of highly
singular generalized functions}

\maketitle

\begin{abstract}
We study the localization properties of generalized
functions defined on a broad class of spaces of entire
analytic test functions. This class, which includes all
Gelfand--Shilov spaces $S^\beta_\alpha(\R^k)$ with
$\beta<1$, provides a convenient language for describing
quantum fields with a highly singular infrared behavior. We
show that the carrier cone notion, which replaces the
support notion, can be correctly defined for the considered
analytic functionals. In particular, we prove that each
functional has a uniquely determined minimal carrier cone.
\end{abstract}


\section{Introduction}

In this paper, we study the localization properties of
generalized functions defined on spaces of entire analytic
test functions. The usual definition of the support of a
generalized function is inapplicable in this case because
of the lack of test functions with compact support (this
difficulty is well known in the theory of hyperfunctions,
where real-analytic test functions are used; see, e.g.,
Chap.~9 in~\cite{Hoermander}). The problem of finding a
reasonable substitute for the support notion is important
for extending the Wightman axiomatic approach to quantum
gauge theory. Because of severe infrared singularities,
gauge fields are generally well defined only under smearing
with entire analytic functions in the momentum space (for
example, this is the case for the Schwinger model in an
arbitrary $\alpha$-gauge~\cite{CapFer}) and can therefore
be treated neither in the original Wightman
framework~\cite{SW} using tempered distributions nor in a
more general framework~\cite{BN} based on Fourier
hyperfunctions. This produces the problem of generalizing
the spectral condition~\cite{MoschStroc}, whose standard
formulation in terms of vacuum expectations relies heavily
on the notion of the support of a generalized function

The localization properties of functionals defined on the
Gelfand-Shilov spaces $S^\beta_\alpha$ with $\beta<1$ were
studied in~\cite{Sol1,Sol2} (see~\cite{GS} for the
definition and properties of $S^\beta_\alpha$; if
$\beta<1$, then $S^\beta_\alpha$ consists of entire
analytic functions). It was shown that a carrier cone
notion, which replaces the support notion, can be
introduced consistently for such functionals. In
particular, it was proved that each element of
$S^{\prime\beta}_{\alpha}(\R^k)$ (the topological dual of
$S^{\beta}_{\alpha}(\R^k)$) has a uniquely determined
minimal carrier cone. Here, we extend the results
in~\cite{Sol1,Sol2} to a broader class of test function
spaces previously used to analyze the spectral properties
of sums of infinite series in Wick powers of
indefinite-metric free fields~\cite{SS}. This class is
defined as follows.

\begin{defin} \label{d1}
Let $\alpha(s)$ and $\beta(s)$ be unbounded continuous
monotonically increasing functions on the semiaxis $s\geq
0$. Let $\beta$ be convex, and let there be a constant
$\varkappa>0$ such that the function
$\alpha(s)/s^\varkappa$ is nondecreasing for sufficiently
large $s$. For any $A,B>0$, $\mathcal E^{\beta,
B}_{\alpha,A}(\R^k)$ denotes the Banach space of all entire
analytic functions on $\oC^k$ with the finite norm
\[
\sup_{z=x+iy\in\oC^k}|f(z)|^{\alpha(|x/A|)- \beta(B|y|)}.
\]
The space $\mathcal E^{\beta}_{\alpha}(\R^k)$ is defined as
the union $\bigcup_{A,B>0} \mathcal E^{\beta,
B}_{\alpha,A}(\R^k)$ endowed with the inductive limit
topology.
\end{defin}

For definiteness, we everywhere assume that the norm
$|\cdot|$ on $\oC^k$ is uniform: $|z|=\max_{1\leq j\leq k}
|z_j|$. For convex $\alpha$, the spaces $\mathcal
E^\beta_\alpha$ coincide with the spaces of type $W$
described in Chap.~1 in~\cite{GS1}. If
$\alpha(s)=s^{1/\mu}$ and $\beta(s)=s^{1/(\nu-1)}$,
$\nu<1$, then $\mathcal
E^\beta_\alpha(\R^k)=S^\nu_\mu(\R^k)$ (to avoid confusion,
we use $S^\nu_\mu$ instead of the standard
$S^\beta_\alpha$). We call a cone $W$ a conic neighborhood
of a cone $U$ if $W$ has an open projection\footnote{By
definition, the projection $\Pr W$ of a cone $W\subset
\R^k$ is the image of $W\setminus\{0\}$ under the canonical
map from $\R^k\setminus\{0\}$ to the sphere $\mathbb
S_{k-1}=(\R^k\setminus\{0\})/\R_+$; the projection of $W$
is assumed to be open in the topology of this sphere. We
note that the degenerate cone $\{0\}$ is a cone with an
open (empty) projection.} and contains $U$. To define
carrier cones, in addition to $\mathcal
E^\beta_\alpha(\R^k)$, we introduce similar spaces
associated with cones in $\R^k$.

\begin{defin} \label{d2}
Let $U$ be a nonempty cone in $\R^k$ and $\alpha$ and
$\beta$ satisfy the conditions in Definition~\ref{d1}. For
any $A,B>0$, $\mathcal E^{\beta, B}_{\alpha,A}(U)$ denotes
the Banach space of all entire analytic functions on
$\oC^k$ with the finite norm
\[
\|f\|_{U,A,B}=\sup_{z\in\oC^k}|f(z)|e^{-\rho_{U,A,B}(z)},
\]
where
\begin{equation}\label{4}
\rho_{U,A,B}(x+iy)=
-\alpha(|x/A|)+\beta(B|y|)+\beta(B\delta_U(x))
\end{equation}
and $\delta_U(x)=\inf_{x'\in U}|x-x'|$ is the distance from
$x$ to $U$. The space $\mathcal E^{\beta}_{\alpha}(U)$ is
defined by the relation $\mathcal
E^{\beta}_{\alpha}(U)=\bigcup_{A,B>0,\,W\supset U} \mathcal
E^{\beta, B}_{\alpha,A}(W)$, where $W$ ranges all conic
neighborhoods of $U$ and the union is endowed with the
inductive limit topology.
\end{defin}

If $U=\R^k$, then Definition~\ref{d2} is equivalent to
Definition~\ref{d1}. Hereafter, we assume that all
considered cones are nonempty. A closed cone $K$ is called
a carrier cone of a functional $u\in \mathcal E^{\prime
\beta}_{\alpha}(\R^k)$ if $u$ has a continuous extension to
the space $\mathcal E_{\alpha}^\beta(K)$. Our main result
in this paper is the following theorem.

\begin{teor}
\label{t1} Let the functions $\alpha$ and $\beta$ satisfy
the conditions in Definition~$\ref{d1}$. If the space
$\mathcal E^\beta_\alpha(\R^k)$ is nontrivial (i.e.,
contains nonzero functions), then the following statements
hold:
\begin{enumerate}
\item[1.] The space $\mathcal E_{\alpha}^\beta(\R^k)$ is dense in
$\mathcal E_{\alpha}^{\beta}(U)$ for any cone
$U\subset\R^k$.
\item[2.] If $K_1$ and $K_2$ are closed cones in
$\R^k$, then for any $u\in \mathcal E^{\prime
\beta}_{\alpha}(\R^k)$ carried by $K_1\cup K_2$, there
exist $u_{1,2}\in \mathcal E^{\prime \beta}_{\alpha}(\R^k)$
carried by $K_{1,2}$ such that $u=u_1+u_2$.
\item[3.] If both $K_1$ and $K_2$ are carrier cones of
$u\in \mathcal E^{\prime \beta}_{\alpha}(\R^k)$, then so is
$K_1\cap K_2$.
\end{enumerate}
\end{teor}

We note that analogous results for Gelfand--Shilov spaces
$S^\nu_\mu$ were proved differently for $\nu=0$ and
$0<\nu<1$ in~\cite{Sol1},~\cite{Sol2}. Our approach here
allows treating both these cases the same.

Statement~1 in Theorem~\ref{t1} shows that the space of the
functionals with the carrier cone $K$ is naturally
identified with the space $\mathcal E_{\alpha}^{\prime
\beta}(K)$. By Definition~\ref{d2}, we have
\[
\mathcal E^\beta_\alpha(K)=\bigcup_{W\supset K} \mathcal
E^{\beta}_{\alpha}(W),
\]
where the union is taken over all conic neighborhoods of
$K$ and is endowed with the inductive limit topology. It
hence follows from Statement~1 in Theorem~\ref{t1} that a
functional $u\in \mathcal E_{\alpha}^{\prime \beta}(\R^k)$
is carried by $K$ if and only if $u$ has a continuous
extension to the space $\mathcal E_{\alpha}^{\beta}(W)$ for
every conic neighborhood $W$ of $K$. Statement~3 in
Theorem~\ref{t1} implies that the intersection of an
arbitrary family $\{K_\omega\}_{\omega\in\Omega}$ of
carrier cones of a functional $u\in \mathcal E^{\prime
\beta}_{\alpha}(\R^k)$ is again a carrier cone of $u$.
Indeed, let $W$ be a conic neighborhood of
$K=\bigcap_{\omega\in\Omega} K_\omega$. Then by standard
compactness arguments (cf.~the proof of statement~A in
Lemma~$\ref{l6}$ below), there exists a finite family
$\omega_1,\ldots,\omega_n\in\Omega$ such that $\tilde
K=\bigcap_{j=1}^n K_{\omega_j}\subset W$. By Statement~3 in
Theorem~\ref{t1}, $\tilde K$ is a carrier cone of $u$, and
$u$ therefore has a continuous extension to $\mathcal
E^{\beta}_{\alpha}(W)$. Hence, $K$ is a carrier cone of
$u$. In particular, each functional $u\in \mathcal
E^{\prime \beta}_{\alpha}(\R^k)$ has a uniquely defined
minimal carrier cone, the intersection of all carrier
cones~of~$u$.

The proof of Theorem~\ref{t1} essentially relies on using
H\"ormander's $L_2$ estimates for the solutions of the
inhomogeneous Cauchy--Riemann equations\footnote{Here and
hereafter, we use the short notation $\bar\partial_j$ for
$\partial/\partial\bar z_j$.} $\bar\partial_j \psi=\eta_j$,
$j=1,\ldots,k$. These estimates ensure the existence of a
solution $\psi$ that is square-integrable with respect to
the weight function $e^{-\rho}/(1+|z|)^{2}$ if the $\eta_j$
are square-integrable with respect to the weight function
$e^{-\rho}$ and $\rho$ is a plurisubharmonic function on
$\oC^k$. To illustrate how this result applies in our case,
we briefly outline the proof of statement~1 in the theorem.

Let $\chi(z)$ be a smooth function on $\oC^k$ vanishing for
large $|z|$ and equal to unity in a neighborhood of the
origin. For $f\in\mathcal E^{\beta}_{\alpha}(U)$, we
construct an approximating sequence by setting
$f_n(z)=f(z)\chi(z/n)-\psi_n(z)$, where the terms $\psi_n$
are introduced to ensure the analyticity of $f_n$. This
latter condition means that $\psi_n$ satisfy the equations
$\bar\partial_j \psi_n(z)=n^{-1}f(z)(\bar\partial_j
\chi)(z/n)$. Hence, we can use the $L_2$ estimates to prove
that the $\psi_n$ can be chosen sufficiently small that
$f_n\in \mathcal E^{\beta}_{\alpha}(\R^k)$ and $f_n\to f$
in $\mathcal E^{\beta}_{\alpha}(U)$. But this strategy
implies using $L_2$-type norms, while $\mathcal
E^{\beta}_{\alpha}(U)$ are defined by supremum norms. We
resolve this problem in Sec.~\ref{s2}, where we derive an
equivalent representation for $\mathcal
E^{\beta}_{\alpha}(U)$ in terms of Hilbert spaces. Another
complication is that the weight functions
$e^{-\rho_{U,A,B}}$ in Definition~\ref{d2} are not
appropriate for $L_2$ estimates, because the functions
$\rho_{U,A,B}$ are not plurisubharmonic. In Sec.~\ref{s3},
we overcome this difficulty by constructing suitable
plurisubharmonic approximations for $\rho_{U,A,B}$. We
prove Theorem~\ref{t1} in Sec.~\ref{s4}.

\section{Hilbert space representation for $\mathcal
E^\beta_\alpha$}\label{s2}

Let $A,B>0$ and $U$ be a cone in $\R^k$. We let
$H^{\beta,B}_{\alpha,A}(U)$ denote the Hilbert space of all
entire functions on $\oC^k$ having the finite norm
\begin{equation}\label{p}
\|f\|'_{U,A,B}=\left[\int
|f(z)|^2e^{-2\rho_{U,A,B}(z)}\,\di\lambda(z)\right]^{1/2},
\end{equation}
where $\di\lambda$ is the Lebesgue measure on $\oC^k$ and
$\rho_{U,A,B}$ is given by~(\ref{4}). We let
$\tilde{\mathcal{E}}^{\beta,B}_{\alpha,A}(U)$ denote the
space
$\bigcap_{A'>A,\,B'>B}\mathcal{E}^{\beta,B'}_{\alpha,A'}(U)$
endowed with the topology defined by the norms
$\|\cdot\|_{U,A',B'}$.

\begin{lem}\label{l0}
Let $A,B>0$, $U$ be a cone in $\R^k$, and $\alpha$ and
$\beta$ satisfy the conditions in Definition~$\ref{d1}$.
Then $\tilde{\mathcal{E}}^{\beta,B}_{\alpha,A}(U)$ is a
nuclear Fr\'echet space coinciding with
$\bigcap_{A'>A,\,B'>B} H^{\beta,B'}_{\alpha,A'}(U)$ both as
a set and topologically.
\end{lem}
\begin{proof}
The space $\tilde{\mathcal{E}}^{\beta,B}_{\alpha,A}(U)$
belongs to the class of the spaces $\mathcal H(M)$
introduced in~\cite{SS1}. The spaces $\mathcal H(M)$ and
$\mathcal H_p(M)$ for $p\geq 1$ are defined\footnote{The
definition of $\mathcal H(M)$ and $\mathcal H_p(M)$ given
here is slightly less general than that in~\cite{SS1} but
suffices for our purposes.} by a family
$M=\{M_\gamma\}_{\gamma\in\Gamma}$ of strictly positive
continuous functions on $\oC^k$ and consist of all entire
analytic functions on $\oC^k$ having the respective finite
norms
\[
\sup_{z\in \oC^k} M_\gamma(z)|f(z)|,\quad
\left[\int(M_\gamma(z))^p|f(z)|^p\,\di\lambda(z)\right]^{1/p}.
\]
We suppose that (a) for any $\gamma_1,\gamma_2\in\Gamma$,
we can find $\gamma\in\Gamma$ and $C>0$ such that
$M_\gamma\geq C (M_{\gamma_1}+M_{\gamma_2})$ and (b) there
exists a countable set $\Gamma'\subset \Gamma$ such that
for every $\gamma\in \Gamma$, we can find $\gamma'\in
\Gamma'$ and $C>0$ such that $C M_\gamma\leq M_{\gamma'}$.
Let $\Gamma=\{(A',B'):A'>A,\,B'>B\}$ and
$M_{A',B'}(z)=e^{-\rho_{U,A',B'}(z)}$. Then all the above
conditions are satisfied, and we have $\mathcal H(M)=
\tilde{\mathcal{E}}^{\beta,B}_{\alpha,A}(U)$ and $\mathcal
H_2(M)=\bigcap_{A',\,B'}H^{\beta,B'}_{\alpha,A'}(U)$. By
Lemma~12 in~\cite{SS1}, $\mathcal H(M)$ is a nuclear
Fr\'echet space coinciding with $\mathcal H_p(M)$ for any
$p\geq 1$ if the following conditions are satisfied:
\begin{itemize}
\item[(I)] For any $\gamma\in \Gamma$, there
exists $\gamma'\in \Gamma$ such that
$M_\gamma(z)/M_{\gamma'}(z)$ is integrable on $\oC^k$ and
tends to zero as $|z|\to\infty$.
\item[(II)] For any $\gamma\in \Gamma$, there exist
$\gamma'\in \Gamma$, a neighborhood of the origin $\mathcal
B$ in $\oC^k$, and $C>0$ such that $M_\gamma(z)\leq
CM_{\gamma'}(z+\zeta)$ for any $z\in \oC^k$ and $\zeta\in
\mathcal B$.
\end{itemize}
In the considered case, the satisfaction of conditions~(I)
and~(II) respectively follows from Lemmas~\ref{l1}
and~\ref{l2} below. The lemma is proved.
\end{proof}

\begin{lem}\label{l1}
Let $U$ be a cone in $\R^k$ and $\alpha$ and $\beta$
satisfy the conditions in Definition~$\ref{d1}$. For any
$A'>A>0$ and $B'>B>0$, $\sigma,\tau>0$ can be found such
that
\begin{equation}\label{5}
\rho_{U,A',B'}(z)-\rho_{U,A,B}(z)+C\geq \sigma
|z|^\tau,\quad z\in\oC^k,
\end{equation}
where $C$ is a constant and $\rho_{U,A,B}$ is given by
$(\ref{4})$.
\end{lem}
\begin{proof}
Without loss of generality, we assume that $\beta(0)=0$.
Let $\varkappa$ satisfy the conditions in
Definition~\ref{d1}, $s_0> 0$ be such that $\alpha(s_0)>0$,
$\beta(s_0)>0$, and the function
$\mu(s)=\alpha(s)/s^\varkappa$ be nondecreasing for $s\geq
s_0$. For $|x|\geq A's_0$, we have
\begin{multline} \nonumber
\alpha\left(\frac{|x|}{A}\right)-
\alpha\left(\frac{|x|}{A'}\right)=
\frac{|x|^\varkappa}{A^\varkappa}\mu\left(\frac{|x|}{A}\right)
- \frac{|x|^\varkappa}{A^{\prime\varkappa}}
\mu\left(\frac{|x|}{A'}\right)\geq \\
\geq\left(\frac{1}{A^\varkappa}-\frac{1}{A^{\prime\varkappa}}\right)
\mu\left(\frac{|x|}{A}\right)|x|^\varkappa \geq
\left(\frac{1}{A^\varkappa}-\frac{1}{A^{\prime\varkappa}}\right)\mu(s_0)|x|^\varkappa.
\end{multline}
Because $\beta(0)=0$, the convexity of $\beta$ implies that
$\beta(s)\leq t\beta(s/t)$ for any $s\geq 0$ and $0<t\leq
1$. It hence follows that $\beta(s)/s$ is a nondecreasing
function. We therefore have
\[
\beta(B'|y|)-\beta(B|y|)\geq
(B'-B)\frac{\beta(B'|y|)}{B'|y|}|y|\geq
(B'-B)\frac{\beta(s_0)}{s_0}|y|
\]
for $|y|\geq s_0/B'$. Setting $\tau=\min(1,\varkappa)$ and
summing the estimates for $\alpha$ and $\beta$, we find
that inequality (\ref{5}) with $C=0$ holds for large $|z|$
if $\sigma$ is sufficiently small. Because all considered
functions are continuous, adding a sufficiently large
positive constant to the left-hand side ensures that the
required bound holds for all $z\in \oC^k$. The lemma is
proved.
\end{proof}

\begin{lem}\label{l2}
Let $R>0$, $U$ be a cone in $\R^k$, $\alpha$ and $\beta$ be
nondecreasing functions on $[0,\infty)$, and $\rho_{U,A,B}$
be given by $(\ref{4})$. For any $A'>A>0$ and $B'>B>0$,
there exists a constant $C$ such that
\[
\rho_{U,A,B}(z+\zeta)\leq \rho_{U,A',B'}(z)+C, \quad
z,\zeta\in\oC^k,\quad |\zeta|\leq R.
\]
\end{lem}
\begin{proof}
Without loss of generality, we can assume that $\alpha$ and
$\beta$ are nonnegative. It then follows from the
monotonicity of $\alpha$ and $\beta$ that
$\alpha((s+R)/A')\leq \alpha(s/A)+\alpha(R/(A'-A))$ and
$\beta(B(s+R))\leq \beta(B's)+\beta(RBB'/(B'-B))$. Let
$z=x+iy$ and $\zeta=\xi+i\eta$ be such that $|\zeta|\leq
R$. Because $\delta_U(x+\xi)\leq \delta_U(x)+|\xi|$, we
have
\begin{align}
&\alpha\left(\frac{|x|}{A'}\right)\leq
\alpha\left(\frac{|x+\xi|+R}{A'}\right)\leq
\alpha\left(\frac{|x+\xi|}{A}\right)+
\alpha\left(\frac{R}{A'-A}\right),\nonumber\\
&\beta(B|y+\eta|)+\beta(B\delta_U(x+\xi))\leq
\beta(B'|y|)+\beta(B'\delta_U(x))+2\beta\left(\frac{RBB'}{B'-B}\right).
\nonumber
\end{align}
Summing these inequalities yields the required estimate.
\end{proof}

\begin{cor} \label{cor1} If $f\in\mathcal
E^{\beta}_\alpha(U)$, then $f(\cdot+\zeta)\in\mathcal
E^{\beta}_\alpha(U)$ for any $\zeta\in\oC^k$.
\end{cor}

We recall that dual Fr\'echet--Schwartz (DFS) spaces are by
definition the inductive limits of sequences of locally
convex spaces with injective compact linking maps
(see~\cite{Komatsu}).

\begin{lem}\label{l3}
Let $U$ be a cone in $\R^k$ and $\alpha$ and $\beta$
satisfy the conditions in Definition~$\ref{d1}$. Then
$\mathcal E^\beta_\alpha(U)$ is a nuclear DFS space
coinciding (both as a set and topologically) with the space
\[
\bigcup_{A,B>0,\,W\supset U}H^{\beta,B}_{\alpha,A}(W),
\]
where $W$ ranges all conic neighborhoods of $U$ and the
union is endowed with the inductive limit topology.
\end{lem}
\begin{proof}
Let $A'>A>0$ and $B'>B>0$, and let $W\supset W'$ be conic
neighborhoods of $U$. Then we have continuous inclusion
maps $\mathcal E^{\beta,B}_{\alpha,A}(W)\to
\tilde{\mathcal{E}}^{\beta,B}_{\alpha,A}(W)\to \mathcal
E^{\beta,B'}_{\alpha,A'}(W')$. We therefore have
\begin{equation}\label{8}
\mathcal E^\beta_\alpha(U)=\bigcup_{A,B>0,\,W\supset
U}\tilde{\mathcal{E}}^{\beta,B}_{\alpha,A}(W).
\end{equation}
Because countable inductive limits of nuclear spaces are
nuclear (see, e.g., the corollary to Theorem~III.7.4
in~\cite{Schaefer}), the nuclearity of $\mathcal
E^\beta_\alpha(U)$ follows from Lemma~\ref{l0}. Because all
continuous maps from nuclear spaces to Banach spaces are
nuclear (Theorem~III.7.2 in~\cite{Schaefer}), the inclusion
map $\mathcal E^{\beta,B}_{\alpha,A}(W)\to \mathcal
E^{\beta,B'}_{\alpha,A'}(W')$ is nuclear as a composition
of a nuclear map and a continuous map. It hence follows
that $\mathcal E^\beta_\alpha(U)$ is a DFS space because
nuclear maps are compact (Corollary~1 to Theorem~III.7.1
in~\cite{Schaefer}). By Lemma~\ref{l0}, we have continuous
inclusions $H^{\beta,B}_{\alpha,A}(W)\to
\tilde{\mathcal{E}}^{\beta,B}_{\alpha,A}(W)\to
H^{\beta,B'}_{\alpha,A'}(W')$. In view of (\ref{8}), this
implies that $\mathcal
E^\beta_\alpha(U)=\bigcup_{A,B>0,\,W\supset
U}H^{\beta,B}_{\alpha,A}(W)$. The lemma is proved.
\end{proof}

\section{Plurisubharmonic approximations}\label{s3}

We recall that the norm $|\cdot|$ is assumed to be uniform.

\begin{teor} \label{pl:t}
Let $A,B>0$, $U$ be a nonempty cone in $\R^k$, $\alpha$ be
a continuous nondecreasing function on $[0,\infty)$, and
$\beta$ be a continuous convex nondecreasing function on
$[0,\infty)$. If there exists an entire function $\varphi$
on $\oC$ which is not identically zero and satisfies the
bound
\begin{equation}\label{pl:000}
|\varphi(z)|\leq e^{\beta(|B_0 y|)-\alpha(|x/A_0|)},\quad
z=x+iy\in\oC,
\end{equation}
for some $A_0,B_0>0$, then for any $R>0$, there exists a
plurisubharmonic function $\rho_R$ on $\oC^k$ such that
\begin{align}
& \rho_R(z)\leq \rho_{\R^k,A',B'}(z)+\beta(2BeR),\quad z=x+iy\in\oC^k,\nonumber\\
&\rho_R(z)\leq \rho_{U,A',B'}(z),\quad z\in\oC^k,\label{pl:6a}\\
&\rho_R(z)\geq \rho_{U,A,B}(z)-H, \quad |x|\leq R,\nonumber
\end{align}
where $\rho_{U,A,B}$ is given by $(\ref{4})$, $H$ is a
constant independent of $R$, $A'=2A$, and
$B'=(2ek+1)B+4kA_0B_0/A$. If $\alpha$ is concave, then we
can set $A'=A$.
\end{teor}

\begin{cor}\label{cor2}
Under the conditions of Theorem~$\ref{pl:t}$, there exists
a plurisubharmonic function $\rho$ such that
\[
\rho_{U,A,B}(z)-H\leq \rho(z)\leq \rho_{U,A',B'}(z),\quad
z\in\oC^k,
\]
where $\rho_{U,A,B}$ is given by $(\ref{4})$, $H$ is a
constant, $A'=2A$, and $B'=(2ek+1)B+4kA_0B_0/A$. If
$\alpha$ is concave, then we can set $A'=A$.
\end{cor}

{\bf Proof.} Let $\rho_R$ satisfy the conditions in
Theorem~\ref{pl:t}. Then the function $\rho(z)=
\varlimsup_{z'\to z}\sup_{R>0}\rho_R(z')$ is
plurisubharmonic (Sec.~II.10.3 in~\cite{V}) and satisfies
the required estimate.

\medskip
To prove Theorem~\ref{pl:t}, we need two lemmas.

\begin{lem} \label{pl:l1}
Let $\alpha$ and $\beta$ be continuous nondecreasing
functions on $[0,\infty)$, and let there exist an entire
analytic function $\varphi$ on $\oC$ that is not
identically zero and satisfies the bound
\begin{equation}\label{pl:1}
|\varphi(z)|\leq e^{\beta(|y|)-\alpha(|x|)},\quad
z=x+iy\in\oC.
\end{equation}
Then there exist a plurisubharmonic function $\rho$ on
$\oC^k$ and a constant $H$ such that
\begin{equation}\label{pl:1a}
-\alpha(2|x|)-k\beta(4|y|)-H\leq \rho(z)\leq
k\beta(4|y|)-\alpha(|x|),\quad z=x+iy\in\oC^k.
\end{equation}
If $\alpha$ is concave, then there exist a plurisubharmonic
function $\rho$ on $\oC^k$ and a constant $H$ such that
\begin{equation}\label{pl:1b}
-\alpha(|x|)-k\beta(2|y|)-H\leq \rho(z)\leq
k\beta(2|y|)-\alpha(|x|),\quad z=x+iy\in\oC^k.
\end{equation}
\end{lem}
\begin{proof}
Without loss of generality, we can assume that
$\alpha(0)=\beta(0)=0$ and $\varphi(0)\neq 0$ (if $\varphi$
has a zero of order $n$ at $z=0$, then we can replace
$\varphi$ with $\tilde \varphi(z)=C\varphi(z)/z^n$; the
function $\tilde \varphi$ satisfies~(\ref{pl:1}) for
sufficiently small $C$). We set
\begin{equation} \label{num10}
\begin{aligned}
&\tilde
\rho(z)=\sup_{\zeta\in\oC^k}\{\Phi(z-\zeta)+M(\zeta)\},\\
&M(\zeta)=\inf_{z'=x'+iy'\in
\oC^k}\{-\Phi(z'-\zeta)+k\beta(4|y'|)-\alpha(|x'|)\},
\end{aligned}
\end{equation}
where $\Phi(z)=\sum_{j=1}^k\log|\varphi(2z_j)|$. We
obviously have $\tilde \rho(z)\leq
k\beta(4|y|)-\alpha(|x|)$. Because $\Phi$ is
plurisubharmonic, $\rho(z)=\varlimsup_{z'\to
z}\tilde\rho(z)$ is also a plurisubharmonic function (see
Sec.~II.10.3 in~\cite{V}). In view of the continuity of
$\alpha$ and $\beta$, we have $\tilde\rho(z)\leq\rho(z)\leq
k\beta(4|y|)-\alpha(|x|)$, and it remains to show that
$\tilde\rho(z)\geq -\alpha(2|x|)-k\beta(4|y|)-H$. It
follows from~(\ref{pl:1}) that
\[
-\Phi(z'-z)\geq \alpha(2|x'-x|)-k\beta(2|y'-y|),\quad
\zeta=\xi+i\eta,
\]
and setting $H=-\Phi(0)=-k\log|\varphi(0)|$, we obtain
\begin{equation}\label{pl:2}
\tilde \rho(z)\geq -H+M(z) \geq \inf_{x',y'\in
\R^k}\{k\beta(4|y'|)-k\beta(2|y'-y|)+\alpha(2|x'-x|)-
\alpha(|x'|)\}-H.
\end{equation}
Because both $\alpha$ and $\beta$ are nonnegative and
monotonic, we have
\begin{equation}\label{pl:aa}
\beta(2|y'|)-\beta(|y'-y|)\geq -\beta(2|y|),\quad
\alpha(2|x'-x|)- \alpha(|x'|)\geq -\alpha(2|x|).
\end{equation}
Substituting these inequalities in~(\ref{pl:2}), we obtain
the required lower estimate for $\tilde\rho$. Thus,
(\ref{pl:1a}) is proved.

Now let $\alpha$ be concave. We replace $\beta(4|y'|)$ with
$\beta(2|y'|)$ in definition~(\ref{num10}) of $M(\zeta)$
and modify $\Phi(z)$ by setting
$\Phi(z)=\sum_{j=1}^k\log|\varphi(z_j)|$. Defining $\tilde
\rho$ and $\rho$ as above, we obtain
$\tilde\rho(z)\leq\rho(z)\leq k\beta(2|y|)-\alpha(|x|)$.
Proceeding as above, we obtain the estimate
\begin{equation}\label{pl:22}
\tilde \rho(z)\geq \inf_{x',y'\in
\R^k}\{k\beta(2|y'|)-k\beta(|y'-y|)+\alpha(|x'-x|)-
\alpha(|x'|)\}-H.
\end{equation}
Because $\alpha$ is concave and $\alpha(0)=0$, we have
$\alpha(s+t)\leq \alpha(s)+\alpha(t)$ for any $s,t\geq 0$.
It hence follows that
\[
\alpha(|x+x'|)\leq \alpha(|x|)+\alpha(|x'|)
\]
for any $x,x'\in\R^k$. Changing $x'\to x'-x$, we obtain
$\alpha(|x'-x|)- \alpha(|x'|)\geq -\alpha(|x|)$.
Substituting this estimate and the first of inequalities
(\ref{pl:aa}) in (\ref{pl:22}) yields $\tilde \rho(z)\geq
-\alpha(|x|)-k\beta(2|y|)-H$, which completes the proof of
(\ref{pl:1b}).
\end{proof}

\begin{lem} \label{pl:l3}
Let $U$ be a cone in $\R^k$. For any $R>0$, there exists a
plurisubharmonic function $\sigma_R$ on $\oC^k$ such that
\begin{align}
& \sigma_R(z)\leq k|y|+R,\quad z=x+iy\in\oC^k,\label{pl:5}\\
&\sigma_R(z)\leq k|y|+\delta_U(x),\quad z\in\oC^k,\label{pl:6}\\
&\sigma_R(z)\geq \delta_U\left(\frac{x}{e}\right), \quad
|x|\leq R,\label{pl:7}
\end{align}
where $\delta_U(x)=\inf_{x'\in U}|x-x'|$ is the distance
from $x$ to $U$.
\end{lem}
\begin{proof}
For any $a>0$, we define the subharmonic function
$\Theta_a$ on $\oC$:
\[
\Theta_a(z)= a\log\left|\frac{\sin (z/a)}{z/a}\right|.
\]
This function satisfies the inequalities
\begin{align}
& \Theta_a(iy)\geq 0,\quad y\in\R, \label{pl:3}\\
& \Theta_a(z)\leq
|y|-a\log^+\left(\frac{|x|}{a}\right),\quad
z=x+iy\in\oC,\label{pl:4}
\end{align}
where $\log^+(r)=\max(\log r,0)$. Indeed, because
\[
\Theta_a(iy)=a\log\left(\frac{\sinh (y/a)}{y/a}\right),
\]
estimate~(\ref{pl:3}) follows from the inequality $\sinh
y/y\geq 1$, $y\in\R$. Further, it follows from the
inequalities
\[
|\sin z|\leq e^{|y|},\quad \left|\frac{\sin
z}{z}\right|\leq e^{|y|},\quad z=x+iy\in\oC,
\]
that
\[
\left|\frac{\sin(z/a)}{z/a}\right|\leq
e^{|y/a|}\min(1,a/|x|).
\]
Passing to the logarithms, we obtain~(\ref{pl:4}). We now
set
\begin{align*}
&\tilde \sigma_R(z)=\sup_{a>0,\,\xi\in\R^k,\,|\xi|\leq R}\{\Phi_a(z-\xi)+M_a(\xi)\},\\
&M_a(\xi)=\inf_{z'=x'+iy'\in
\oC^k}\{-\Phi_a(z'-\xi)+k|y'|+\delta_U(x')\},
\end{align*}
where $\Phi_a(z)=\sum_{j=1}^k \Theta_a(z_j)$. We obviously
have $\tilde \sigma_R(z)\leq k|y|+\delta_U(x)$. By
inequality~(\ref{pl:4}), $\Phi_a(z-\xi)\leq k|y|$ and
therefore $\tilde \sigma_R(z)\leq
k|y|+\sup_{a>0,\,|\xi|\leq R} M_a(\xi)$. Because
$\Phi_a(0)=0$, it follows from the definition of $M_a$ that
$M_a(\xi)\leq \delta_U(\xi)$. Hence, $\tilde
\sigma_R(z)\leq k|y|+R$. Because $\Phi_a$ are
plurisubharmonic functions, $\sigma_R(z)=\varlimsup_{z'\to
z}\tilde\sigma_R(z)$ is also a plurisubharmonic function,
and it follows from the continuity of $\delta_U(x)$ and
$|y|$ that $\sigma_R$ satisfies~(\ref{pl:5})
and~(\ref{pl:6}). Estimate~(\ref{pl:3}) implies that
$\Phi_a(iy)\geq 0$, $y\in \R^k$. Therefore,
\begin{equation}\label{pl:8}
\tilde\sigma_R(z)\geq \sup_{a>0}(\Phi_a(iy)+M_a(x))\geq
\sup_{a>0}M_a(x),\quad |x|\leq R.
\end{equation}
Using the elementary inequalities
$\sum_{j=1}^k\log^+|x_j|\geq \log^+(|x|)$ and
$\sum_{j=1}^k|y_j|\leq k|y|$, we obtain
\begin{equation}\label{pl:9}
M_a(x)\geq \inf_{x'\in\R^k}\left\{a\log^+
\left(\frac{|x'|}{a}\right)+\delta_U(x+x')\right\}
\end{equation}
from estimate~(\ref{pl:4}). Estimating $\delta_U(x+x')$
from below by $\max(\delta_U(x)-|x'|,0)$ and calculating
the infimum with respect to $x'$, we obtain $M_a(x)\geq
a\log^+(\delta_U(x)/a)$. Let $\delta_U(x)>0$ and
$a_0=\delta_U(x)/e$. In view of (\ref{pl:8}), we find that
\begin{equation}\label{pl:10}
\tilde \sigma_R(z)\geq M_{a_0}(x)\geq \delta_U(x)/e,\quad
|x|\leq R.
\end{equation}
If $\delta_U(x)=0$ and $|x|\leq R$, then the estimate
$\tilde\sigma_R(z)\geq\delta_U(x)/e$ also holds because in
view of (\ref{pl:8}) and (\ref{pl:9}), we have
$\tilde\sigma_R(z)\geq 0$. Hence, (\ref{pl:7}) follows
because $\sigma_R\geq \tilde\sigma_R$. The lemma is proved.
\end{proof}

{\bf Proof of Theorem~$\ref{pl:t}$.} Without loss of
generality, we assume that $\beta(0)=0$. We set
$\rho_R^\prime(z)=\beta(Be\sigma_R(z))$, where $\sigma_R$
is a plurisubharmonic function satisfying the conditions in
Lemma~\ref{pl:l3}. Because a composition of a nondecreasing
convex function with a plurisubharmonic function is
plurisubharmonic (Theorem~4.1.13 and Sec.~4.1 in
\cite{Hoer}), $\rho^\prime_R$ is a plurisubharmonic
function. Because $\beta$ is monotonic, inequalities
(\ref{pl:5})-(\ref{pl:7}) imply the estimates
\begin{align}
& \rho^\prime_R(z)\leq \beta(2Bek|y|)+\beta(2BeR),\quad z=x+iy\in\oC^k,\nonumber\\
&\rho^\prime_R(z)\leq \beta(2Bek|y|)+\beta(2Be\delta_U(x)),\quad z\in\oC^k,\label{pl:61}\\
&\rho^\prime_R(z)\geq \beta(\delta_U(Bx)), \quad |x|\leq
R.\nonumber
\end{align}
By Lemma~\ref{pl:l1}, there exist a plurisubharmonic
function $\rho''$ and a constant $H$ such that
\begin{equation}\label{pl:13}
-\alpha(|x/A|)-k\beta(D|y|)-H\leq \rho''(z)\leq
k\beta(D|y|)-\alpha(|x/A'|),
\end{equation}
where $A'=2A$ and $D=2A_0B_0/A$ ($A'=A$ if $\alpha$ is
concave). We set
$\rho_R(z)=\rho^\prime_R(z)+\rho''(z)+k\beta(D|y|)+\beta(|By|)$.
The function $\rho_R$ is plurisubharmonic because
$\beta(D|y|)$ and $\beta(|By|)$ are convex and are
therefore plurisubharmonic functions. Estimates
(\ref{pl:6a}) with $B'=2kD+(2ek+1)B$ easily follow
from~(\ref{pl:61}), (\ref{pl:13}), and the inequality
\[
2k\beta(D|y|)+\beta(B|y|)+\beta(2Bek|y|)\leq \beta(B'|y|),
\]
which follows from the convexity of $\beta$ and the
condition $\beta(0)=0$.

\section{Proof of Theorem~\ref{t1}}\label{s4}

As above, we let $\di\lambda$ denote the Lebesgue measure
on $\oC^k$. The proof of Theorem~\ref{t1} is based on the
following statement, which is a particular case of
Theorem~4.2.6 in~\cite{Hoer}.

\begin{lem}\label{l4}
Let $\rho$ be a plurisubharmonic function on $\oC^k$ and
$\eta_j$, $j=1,\ldots,k$, be locally square-integrable
functions on $\oC^k$. If
\[
\int |\eta_j(z)|^2
e^{-\rho(z)}\,\di\lambda(z)<\infty
\]
for all $j$ and $\eta_j$ (as generalized functions) satisfy
the compatibility conditions
$\bar\partial_j\eta_l=\bar\partial_l\eta_j$, then the
inhomogeneous Cauchy--Riemann equations
$\bar\partial_j\psi=\eta_j$ have a locally
square-integrable solution satisfying the
estimate\footnote{The estimate in Lemma~\ref{l4} differs
from the estimate in~\cite{Hoer} by the factor $k^2$ in the
right-hand side, which appears because we use the uniform
norm instead of the Euclidean norm used in~\cite{Hoer}.}
\[
2\int |\psi(z)|^2e^{-\rho(z)}(1+|z|^2)^{-2}\,\di\lambda(z)
\leq k^2\sum_{j=1}^k \int |\eta_j(z)|^2
e^{-\rho(z)}\,\di\lambda(z).
\]
\end{lem}

Let $\rho$ be a measurable locally bounded function on
$\oC^k$. We let $L_2(\oC^k,e^{-\rho}\di\lambda)$ denote the
Hilbert space of functions square-integrable with respect
to the measure $e^{-\rho}\di\lambda$ and $H_\rho$ denote
the closed subspace of $L_2(\oC^k,e^{-\rho}\di\lambda)$
consisting of entire analytic functions.

\begin{lem}\label{l5}
Let $\rho_0$, $\rho$, and $\rho'$ be measurable locally
bounded functions on $\oC^k$ such that $\rho_0\leq \rho'$
and $\rho\leq\rho'$. If there exists a plurisubharmonic
function $\rho_R$ for any $R>0$ such that
\begin{align}
& \rho_R(z)+2\log(1+|z|^2)\leq \rho'(z),\quad z\in\oC^k,\label{d:5}\\
&\rho_R(z)+2\log(1+|z|^2)\leq \rho_0(z)+C_R,\quad z\in\oC^k,\label{d:6}\\
&\rho_R(z)\geq \rho(z), \quad |z|\leq R,\label{d:7}
\end{align}
where $C_R$ is a constant, then $H_\rho$ is contained in
the closure of $H_{\rho_0}$ in $H_{\rho'}$.
\end{lem}
\begin{proof}
Let $f\in H_\rho$ and $\chi$ be a smooth function on
$\oC^k$ such that $0\leq\chi\leq 1$, $\chi(z)=1$ for
$|z|\leq 1$, and $\chi(z)=0$ for $|z|\geq 2$. We set
$g_n(z)=f(z)\chi(z/n)$. Because $\bar\partial_j
g_n(z)=n^{-1}f(z)(\bar\partial_j \chi)(z/n)$ vanishes for
$|z|\geq 2n$, it follows from (\ref{d:7}) that
\[
\int |\bar\partial_j g_n(z)|^2
e^{-\rho_{2n}(z)}\,\di\lambda(z)\leq \int |\bar\partial_j
g_n(z)|^2 e^{-\rho(z)}\,\di\lambda(z)\leq
\frac{a}{n^2}\|f\|^2_\rho,\quad j=1,2,\ldots,k,
\]
where $\|\cdot\|_\rho$ is the norm in
$L_2(\oC^k,e^{-\rho}\di\lambda)$ and
$a=\sup_{z,j}|\bar\partial_j\chi(z)|^2$. By Lemma~\ref{l4},
there exists a locally square-integrable function $\psi_n$
on $\oC^k$ such that $\bar\partial_j\psi_n=\bar\partial_j
g_n$ and
\begin{equation}\label{d:8}
\int |\psi_n|^2
e^{-\rho_{2n}(z)}(1+|z|^2)^{-2}\,\di\lambda(z) \leq
\frac{k^3a}{2n^2}\|f\|_\rho^2.
\end{equation}
In view of (\ref{d:6}), this implies that
$\|\psi_n\|_{\rho_0}<\infty$. Further, we have
$\bar\partial_j (g_n-\psi_n)=0$, and $g_n-\psi_n$ therefore
coincides almost everywhere with an entire analytic
function $f_n$. Because $g_n$ is a function with compact
support, we have $\|g_n\|_{\rho_0}<\infty$ and hence
$f_n\in H_{\rho_0}$. Because $\rho\leq \rho'$, we have
\[
\|f-g_n\|^2_{\rho'}\leq \int_{|z|\geq n} |f(z)|^2
e^{-\rho(z)}\,\di\lambda(z).
\]
Hence, $g_n\to f$ in $L_2(\oC^k,e^{-\rho'}\di\lambda)$. By
(\ref{d:5}) and (\ref{d:8}), we have
$\|\psi_n\|^2_{\rho'}\leq k^3a\|f\|^2_\rho/(2n^2)$.
Therefore, $f_n\to f$ in $H_{\rho'}$, and the lemma is
proved.
\end{proof}

Lemma~\ref{l5} was proved differently in~\cite{Sol2} under
the additional assumption that the $\rho_R$ are smooth. The
simple proof given above is closer to the line of reasoning
sketched in Sec.~5 in~\cite{Sol1}.

We note that the spaces $H^{\beta,B}_{\alpha,A}(U)$
considered in Sec.~\ref{s2} coincide with $H_\rho$ for
$\rho=2\rho_{U,A,B}$.
\par\medskip
{\bf Proof of Theorem~$\ref{t1}$.}
\par
1. Let $f\in\mathcal E^{\beta}_\alpha(U)$. By
Lemma~\ref{l3}, there exist $A,B>0$ and a conic
neighborhood $W$ of $U$ such that $f\in
H^{\beta,B}_{\alpha,A}(W)$. In view of
Corollary~\ref{cor1}, the nontriviality of $\mathcal
E^{\beta}_\alpha(\R^k)$ implies the existence of $A_0,
B_0>0$, and $f_0\in \mathcal
E^{\beta,B_0}_{\alpha,A_0}(\R^k)$ such that $f_0(0)\ne 0$
and $\|f_0\|_{\R^k,A_0,B_0}\leq 1$. Then the entire
function $\varphi(z)=f_0(z,0,\ldots,0)$ on $\oC$ is not
identically zero and satisfies~(\ref{pl:000}). It follows
from Lemma~\ref{l1} and Theorem~\ref{pl:t} that the
functions $\rho=2\rho_{W,A,B}-H$,
$\rho_0=2\rho_{\R^k,A',B'}$, and $\rho'=2\rho_{W,A',B'}$
satisfy the conditions in Lemma~\ref{l5} if $A'>2A$,
$B'>(2ek+1)B+4kA_0B_0/A$, and the constant $H$ is
sufficiently large. By Lemma~\ref{l5}, there exists a
sequence $f_n\in H^{\beta,B'}_{\alpha,A'}(\R^k)$ tending to
$f$ in $H^{\beta,B'}_{\alpha,A'}(W)$. By Lemma~\ref{l3},
the Hilbert topology of $H^{\beta,B'}_{\alpha,A'}(W)$ is
stronger than the topology induced from $\mathcal
E^{\beta}_\alpha(U)$. Hence, $f_n\to f$ in $\mathcal
E^{\beta}_\alpha(U)$.
\par
2. Let $l\colon \mathcal E^{\beta}_\alpha(K_1\cup K_2)\to
\mathcal E^{\beta}_\alpha(K_1)\oplus \mathcal
E^{\beta}_\alpha(K_2)$ and $m\colon \mathcal
E^{\beta}_\alpha(K_1)\oplus \mathcal E^{\beta}_\alpha(K_2)
\to \mathcal E^{\beta}_\alpha(K_1\cap K_2)$ be the
continuous linear maps respectively taking $f$ to $(f,f)$
and $(f_1,f_2)$ to $f_1-f_2$. The map $l$ has a closed
image because we have $\mathcal E^{\beta}_\alpha(K_1)\cap
\mathcal E^{\beta}_\alpha(K_2) = \mathcal
E^{\beta}_\alpha(K_1\cup K_2)$ by Definition~\ref{d2}, and
therefore $\mathrm{Im}\,l =\mathrm{Ker}\, m$. In view of
Lemma~\ref{l3}, this implies that the space
$\mathrm{Im}\,l$ is a DFS space.\footnote{We recall that
the direct sum of a finite family of DFS spaces and a
closed subspace of a DFS space are again DFS spaces
(see~\cite{Komatsu}).} Let $u\in \mathcal
E^{\prime\beta}_\alpha(\R^k)$ be a functional carried by
$K_1\cup K_2$ and $\hat u$ be its continuous extension to
$\mathcal E^{\beta}_\alpha(K_1\cup K_2)$. The linear
functional $\hat ul^{-1}$ is continuous on $\mathrm{Im}\,l$
by the open map theorem (see Theorem~IV.8.3
in~\cite{Schaefer}; it is applicable because DFS spaces as
strong duals of reflexive Fr\'echet spaces are
B-complete~\cite{Komatsu}); by the Hahn--Banach theorem,
there exists a continuous extension $v$ of this functional
to the entire space $\mathcal E^{\beta}_\alpha(K_1)\oplus
\mathcal E^{\beta}_\alpha(K_2)$. Let $v_1$ and $v_2$ be the
respective restrictions of $v$ to $\mathcal
E^{\beta}_\alpha(K_1)$ and $\mathcal
E^{\beta}_\alpha(K_2)$. Then for any $f\in \mathcal
E^{\beta}_\alpha(K_1\cup K_2)$, we have $\hat
u(f)=v(f,f)=v_1(f)+v_2(f)$. This means that $u=u_1+u_2$,
where $u_{1,2}$ are the restrictions of $v_{1,2}$ to
$\mathcal E^{\beta}_\alpha(\R^k)$. By construction,
$u_{1,2}$ are carried by the cones $K_{1,2}$.
\par
3. Let $l$ and $m$ be as defined above, $u\in \mathcal
E^{\prime\beta}_\alpha(\R^k)$ be a functional carried by
both $K_1$ and $K_2$, and $u_{1,2}$ be its continuous
extensions to $\mathcal E^{\prime\beta}_\alpha(K_{1,2})$.
If the map $m$ is surjective, then the open map theorem
implies that $\mathcal E^{\beta}_\alpha(K_1\cap K_2)$ is
topologically isomorphic to the quotient space $(\mathcal
E^{\beta}_\alpha(K_1)\oplus \mathcal
E^{\beta}_\alpha(K_2))/\mathrm{Ker}\,m$. We define the
continuous linear functional $v$ on $\mathcal
E^{\beta}_\alpha(K_1)\oplus \mathcal E^{\beta}_\alpha(K_2)$
by the relation $v(f_1,f_2)=u_1(f_1)-u_2(f_2)$. By
statement~1 in the theorem, $u_1$ and $u_2$ coincide on
$\mathcal E^{\beta}_\alpha(K_1\cup K_2)$, and therefore
$\mathrm{Ker}\,v\supset \mathrm{Im}\,l$. Because
$\mathrm{Ker}\,m = \mathrm{Im}\,l$, this inclusion implies
the existence of a functional $\hat u\in \mathcal
E^{\prime\beta}_\alpha(K_1\cap K_2)$ such that $v=\hat u
m$. If $f_{1,2}\in \mathcal E^{\beta}_\alpha(K_{1,2})$,
then we have $\hat u(f_1)=v(f_1,0)=u_1(f_1)$ and $\hat
u(f_2)=v(0,-f_2)=u_2(f_2)$. Hence, $\hat u$ is a continuous
extension of $u$ to $\mathcal E^{\beta}_\alpha(K_1\cap
K_2)$. Proving statement~3 thus reduces to proving that $m$
is surjective. The latter is implied by the following
result on the decomposition of test functions.

\begin{teor}\label{t2}
Let $\mathcal E^\beta_\alpha(\R^k)$ be nontrivial, $K_1$
and $K_2$ be closed cones in $\R^k$, and $f\in\mathcal
E^\beta_\alpha(K_1\cap K_2)$. Then there exist $f_{1,2}\in
\mathcal E^\beta_\alpha(K_{1,2})$ such that $f=f_1+f_2$.
\end{teor}

In the next lemma, we summarize some simple facts about
cones in $\R^k$ needed for proving Theorem~\ref{t2}.

\begin{lem}\label{l6}
Let $K_1$ and $K_2$ be closed cones in $\R^k$.
\begin{itemize}

\item[{\rm A.}]
For any conic neighborhood $W$ of $K_1\cap K_2$, there
exist conic neighborhoods $V_{1,2}$ of $K_{1,2}$ such that
$\bar V_1\cap \bar V_2\subset W$ (the bar means closure).

\item[{\rm B.}] If $K_1\cap K_2=\{0\}$, then there exists
$\theta>0$ such that $\delta_{K_1}(x)\geq \theta|x|$ for
any $x\in K_2$.
\end{itemize}
\end{lem}
\begin{proof}
A. We let $\mathcal C$ denote the set of all cones in
$\R^k$ containing the origin. By assumption, $K_{1}$,
$K_2$, and $W$ belong to $\mathcal C$. It is easy to see
that the map $U\to \Pr U$ is a bijection between $\mathcal
C$ and the set of all subsets of the sphere $\mathbb
S_{k-1}=(\R^k\setminus\{0\})/\R_+$. Let $Q$ denote its
inverse map. It can be easily verified that both $\Pr$ and
$Q$ preserve closures, unions, and intersections. Hence,
the $\Pr K_{1,2}$ are closed, and we have $\Pr K_1\cap \Pr
K_2\subset \Pr W$. Because $\mathbb S_{k-1}$ is compact,
there exist open neighborhoods $O_{1,2}$ of $\Pr K_{1,2}$
in $\mathbb S_{k-1}$ such that $\bar O_1\cap \bar
O_2\subset \Pr W$. We set $V_{1,2}=Q(O_{1,2})$. Then $\bar
V_1\cap \bar V_2=Q(\bar O_1\cap \bar O_2)\subset Q(\Pr
W)=W$.

B. Let $K_2\ne \{0\}$ (if $K_2=\{0\}$, then the statement
holds for any $\theta>0$). We set $F=\{x\in\R^k:x\in K_2
\mbox{ and }|x|=1\}$ and $\theta=\inf_{x\in
F}\delta_{K_1}(x)$. Because $F$ is compact and $F\cap
K_1=\varnothing$, we have $\theta>0$. It remains to note
that $\delta_{K_1}(x)=|x|\delta_{K_1}(x/|x|)\geq \theta
|x|$ for any nonzero $x\in K_2$.
\end{proof}

\begin{lem}\label{l7}
Let $A,B>0$, and let $U_1$, $U_2$, and $U$ be cones in
$\R^k$ such that $\bar U_1\cap \bar U_2=\{0\}$. If
$\mathcal E^\beta_\alpha(\R^k)$ is nontrivial, then for any
$f\in H^{\beta,B}_{\alpha,A}(U)$, there exist $A',B'>0$ and
$f_{1,2}\in H^{\beta,B'}_{\alpha,A'}(U\cup U_{1,2})$ such
that $f=f_1+f_2$.
\end{lem}
\begin{proof}
There exist conic neighborhoods $V_{1,2}$ of $U_{1,2}$ and
measurable cones $W_{1,2}$ such that
\begin{equation}\label{aaa}
W_1\cup W_2=\R^k,\quad W_1\cap W_2=\{0\},\quad \bar
V_\nu\cap \bar W_\nu =\{0\},\quad \nu=1,2.
\end{equation}
Indeed, applying statement~A in Lemma~$\ref{l6}$ to the
closed cones $\bar U_1$ and $\bar U_2$,\footnote{We note
that the degenerate cone $\{0\}$ is a conic neighborhood of
itself.} we find conic neighborhoods $V_{1,2}$ of $\bar
U_{1,2}$ such that $\bar V_1\cap\bar V_2=\{0\}$. Applying
statement~A in Lemma~$\ref{l6}$ to $\bar V_{1,2}$ again, we
see that there exists a conic neighborhood $W_2$ of $\bar
V_1$ such that $\bar V_2\cap\bar W_2=\{0\}$. We set
$W_1=(\R^k\setminus W_2)\cup\{0\}$. Then the first two
relations in~(\ref{aaa}) obviously hold, and we have $\bar
V_1\cap \bar W_1=\bar V_1\cap W_1=\{0\}$ because $W_1$ is
closed.

Let $g_0$ be a nonnegative smooth function on $\R^k$ such
that $g_0(x)=0$ for $|x|\geq 1$ and
\[
\int_{\R^k}g_0(x)\,\di x=1.
\]
We define smooth functions $g_1$ and $g_2$ on $\oC^k$ by
the relations
\[
g_{\nu}(x+iy)=\int_{W_{\nu}}g_0(x-\xi)\,\di \xi,\quad
x,y\in \R^k,\,\,\,\nu=1,2.
\]
By (\ref{aaa}), we have $g_1+g_2=1$. Applying statement~B
in Lemma~$\ref{l6}$ to the closed cones $\bar U_\nu$ and
$(\R^k\setminus V_\nu)\cup\{0\}$, we conclude that there
exists $\theta\in (0,1)$ such that $\delta_{U_\nu}(x)\geq
\theta |x|$ for $x\notin V_\nu$, $\nu=1,2$. Because
$\delta_U(x)\leq |x|$ for any $x\in\R^k$, we have
\begin{equation}\label{aaa1}
\delta_U(\theta x)\leq \min(\delta_U(x),\theta|x|)\leq
\min(\delta_U(x),\delta_{U_\nu}(x))=\delta_{U\cup
U_\nu}(x), \quad x\notin V_\nu.
\end{equation}
Let $\tilde W_\nu=\{x\in\R^k:\delta_{W_\nu}(x)\leq 1\}$,
$\nu=1,2$. It follows from (\ref{aaa}) and statement~B in
Lemma~$\ref{l6}$ that there exists $\theta'>0$ such that
$\delta_{W_\nu}(x)\geq \theta'|x|$ for $x\in \bar V_\nu$,
$\nu=1,2$. Hence, $\delta_{W_\nu}(x)>1$ for all $x\in
V_\nu$ such that $|x|\geq 1/\theta'$, i.e., the sets
$V_\nu\cap \tilde W_\nu$ are bounded in $\R^k$. In view of
(\ref{aaa1}), this implies that
\begin{equation}\label{aaa2}
\delta_U(x)\leq \delta_{U\cup
U_\nu}\left(\frac{x}{\theta}\right)+C,\quad x\in\tilde
W_\nu,\,\,\nu=1,2,
\end{equation}
where $C$ is a constant. It hence follows that
\begin{equation}\label{aaa3}
\delta_U(x)\leq \delta_{U\cup U_1\cup
U_2}\left(\frac{x}{\theta}\right)+C,\quad x\in\tilde
W_1\cap \tilde W_2.
\end{equation}
Let $\tilde f_{1,2}=f g_{1,2}$. Because $f$ is analytic, we
have $\bar\partial_j \tilde f_1=f\bar\partial_j g_1$,
$j=1,\ldots,k$. By the definition of $g_\nu$, we have
$\mathrm{supp}\,g_\nu\subset\tilde W_\nu$, $\nu=1,2$.
Because $g_1+g_2=1$, this implies
$\mathrm{supp}\,\bar\partial_j g_1\subset\tilde
W_1\cap\tilde W_2$, and in view of (\ref{p}), (\ref{aaa2}),
and (\ref{aaa3}), we obtain
\begin{equation}\label{aaa4}
\|\tilde f_\nu\|'_{U\cup U_\nu,A,\tilde B}\leq \tilde
C\|f\|'_{U,A,B},\quad \|\bar\partial_j\tilde f_1\|'_{U\cup
U_1\cup U_2 ,A,\tilde B}\leq \tilde C\|f\|'_{U,A,B},\quad
\nu=1,2,
\end{equation}
where $j=1,\ldots,k$, $\tilde B=B/\theta$, and $\tilde C$
is a positive constant. As shown in the proof of
statement~1 in Theorem~\ref{t1}, the nontriviality of
$\mathcal E^\beta_\alpha(\R^k)$ implies the existence of an
entire function $\varphi$ on $\oC$
satisfying~(\ref{pl:000}). By Lemma~\ref{l1} and
Corollary~\ref{cor2}, there exist $A'\geq A$, $B'\geq
\tilde B$, and a plurisubharmonic function $\rho$ such that
\begin{equation}\label{aaa5}
\rho_{U\cup U_1\cup U_2,A,\tilde B}(z)-H\leq \rho(z) \leq
\rho_{U\cup U_1\cup U_2,A',B'}(z)- \log(1+|z|^2),\quad
z\in\oC^k,
\end{equation}
where $H$ is a constant. It follows from (\ref{aaa4}) and
(\ref{aaa5}) that
\[
\int |\bar\partial_j\tilde f_1(z)|^2
e^{-2\rho(z)}\,\di\lambda(z)<\infty.
\]
By Lemma~\ref{l4},
the inhomogeneous Cauchy--Riemann equations
$\bar\partial_j\psi=\bar\partial_j\tilde f_1$ have a
locally square-integrable solution such that
\begin{equation}\label{aaa6}
\int |\psi(z)|^2e^{-2\rho(z)}(1+|z|^2)^{-2}\,\di\lambda(z)
<\infty.
\end{equation}
We have $\bar\partial_j(\tilde
f_1-\psi)=\bar\partial_j(\tilde f_2+\psi)=0$; therefore,
there exist entire analytic functions $f_1$ and $f_2$ that
respectively coincide almost everywhere with $\tilde
f_1-\psi$ and $\tilde f_2+\psi$. It follows from the second
inequality in~(\ref{aaa5}) and condition~(\ref{aaa6}) that
$\|\psi\|'_{U\cup U_1\cup U_2, A',B'}<\infty$. In view of
(\ref{aaa4}), it follows that $f_\nu\in
H^{\beta,B'}_{\alpha,A'}(U\cup U_\nu)$, $\nu=1,2$. To
complete the proof, it remains to note that $f=f_1+f_2$
because continuous functions coinciding almost everywhere
are equal.
\end{proof}

{\bf Proof of Theorem~$\ref{t2}$.} By Lemma~\ref{l3}, there
exist $A,B>0$ and a conic neighborhood $W$ of $K_1\cap K_2$
such that $f\in H^{\beta,B}_{\alpha,A}(W)$. By
statement~$A$ in Lemma~$\ref{l6}$, we can find conic
neighborhoods $V_{1,2}$ of $K_{1,2}$ such that $\bar
V_1\cap \bar V_2\subset W$. Because $W$ has an open
projection, the cone $V=(\R^k\setminus W)\cup\{0\}$ is
closed. Applying Lemma~\ref{l7} to the closed cones
$U_{1,2}=\bar V_{1,2}\cap V$ (obviously, $U_1\cap
U_2=\{0\}$), we find $A',B'>0$ and $f_{1,2}\in
H^{\beta,B'}_{\alpha,A'}(W\cup U_{1,2})$ such that
$f=f_1+f_2$. Because $W\cup U_{1,2}\supset V_{1,2}$, it
follows from Lemma~\ref{l3} that $f_{1,2}\in \mathcal
E^{\beta}_{\alpha}(K_{1,2})$. This completes the proof of
Theorem~\ref{t2} and statement~3 in Theorem~\ref{t1}.

\subsection*{Acknowledgements}
The research was supported by the Russian Foundation for
Basic Research (Grant No.~05-01-01049), INTAS (Grant
No.~03-51-6346), the Program for Supporting Leading
Scientific Schools (Grant No.~NSh-4401.2006.2), and the
President of the Russian Federation (Grant
No.~MK-1315.2006.1).

\end{document}